\documentstyle[psfig]{mn}
\input epsf

\title[X-Ray Wakes in Abell~160]
      {X-Ray Wakes in Abell~160}

\author[N. Drake et al.]
  {Nick Drake,$^{1}$ Michael R. Merrifield,$^{2}$ Irini Sakelliou$^{3}$ 
and Jason C. Pinkney$^{4}$ \\ 
   $^{1}$Department of Physics and Astronomy, University of Southampton,
   Highfield, Southampton SO17 1BJ \\ 
   $^{2}$School of Physics and Astronomy, University Park, University of 
   Nottingham, Nottingham NG7 2RD \\ 
   $^{3}$Mullard Space Science Laboratory, University College 
   London, Holmbury St. Mary, Dorking, Surrey RH5 6NT \\ 
   $^{4}$Department of Astronomy, University of Michigan, 
   Ann Arbor, MI~48109--1090, USA}

\date{Accepted . Received ; 
      in original form }

\begin{document}

\maketitle

% *** Abstract ***
\begin{abstract} 

`Wakes' of X-ray emission have now been detected trailing behind a few (at
least seven) elliptical galaxies in clusters.  To quantify how widespread
this phenomenon is, and what its nature might be, we have obtained a deep
(70~ksec) X-ray image of the poor cluster Abell~160 using the
\emph{ROSAT}~HRI.  Combining the X-ray data with optical positions of
confirmed cluster members, and applying a statistic designed to search for
wake-like excesses, we confirm that this phenomenon is observed in
galaxies in this cluster.  The probability that the detections arise from
chance is less than $3.8\times10^{-3}$.  Further, the wakes are not
randomly distributed in direction, but are preferentially oriented
pointing away from the cluster centre.  This arrangement can be explained
by a simple model in which wakes arise from the stripping of their host
galaxies' interstellar media due to ram pressure against the intracluster
medium through which they travel. 

\end{abstract} 


% *** Section One ***
\section{Introduction}
\label{sec:introduction}

With the advent of satellite-based X-ray astronomy, it was discovered that
elliptical galaxies can contain as much interstellar gas as their spiral
kin [for example, see Forman et al.\ (1979)].  In the case of ellipticals,
however, this interstellar medium (ISM) predominantly takes the form of a
hot plasma at a temperature of $\sim 10^7\,{\rm K}$.  The vast majority of
elliptical galaxies are found in clusters, which themselves are permeated
by very hot gas --- the intracluster medium (ICM) --- at a temperature of
$\sim 10^8\,{\rm K}$.  The existence of these two gaseous phases raises
the question of how they interact with each other.  The collisional nature
of such material means that one might expect ram pressure to strip the ISM
from cluster members.  But since the ISM is continually replenished by
mass loss from stellar winds, planetary nebulae, and supernovae, it is not
evident \emph{a priori} that galaxies will be entirely denuded of gas by
this process. 

Observing the X-ray emission from individual cluster galaxies is quite
challenging, since they are viewed against the bright background of the
surrounding ICM.  Sakelliou \& Merrifield (1998) used a deep \emph{ROSAT}
observation to detect the X-ray emission from galaxies in the
moderately-rich cluster Abell~2634.  They showed that the level of galaxy
emission is consistent with the expected X-ray binary content of the
galaxies, and hence that there is no evidence of surviving ISM in this
rich environment. 

When one looks in somewhat poorer environments, one does see evidence for
surviving interstellar gas, and for the stripping process itself. The
best-documented example is M86 in the Virgo Cluster which, when mapped in
X-rays, reveals a tail or plume of hot gas apparently being stripped from
the galaxy by ram pressure [Rangarajan et al.\ (1995), and references
therein].  A similar process appears to be happening to NGC~1404 in the
Fornax Cluster, which displays a clear wake of X-ray emission pointing
away from the cluster centre (Jones et al.\ 1997).  Evidence for a
`cooling wake' formed by gravitational accretion is found in the NGC~5044
group where a soft, linear X-ray feature is seen trailing from NGC~5044
itself (David et al. 1994). 

Given the difficulty of detecting such faint wakes against the bright
background of the ICM, it is quite possible that this phenomenon is
widespread amongst galaxies in poor clusters.  This possibility is
intriguing, since wakes indicate the direction of motion of galaxies on
the plane of the sky, and it has been shown that this information can be
combined with radial velocity data to solve for both the distribution of
galaxy orbits in a cluster and the form of the gravitational potential
(Merrifield 1998).  The existing isolated examples do not tell us,
however, how common wake formation might be amongst cluster galaxies:
although the observed wakes might represent the most blatant examples of
widespread on-going ISM stripping, the galaxies in question might have
merged with their current host clusters only recently, or be in some other
way exceptional.

% * Figure 1 *
\begin{figure*}
 \begin{center}
 \psfig{file=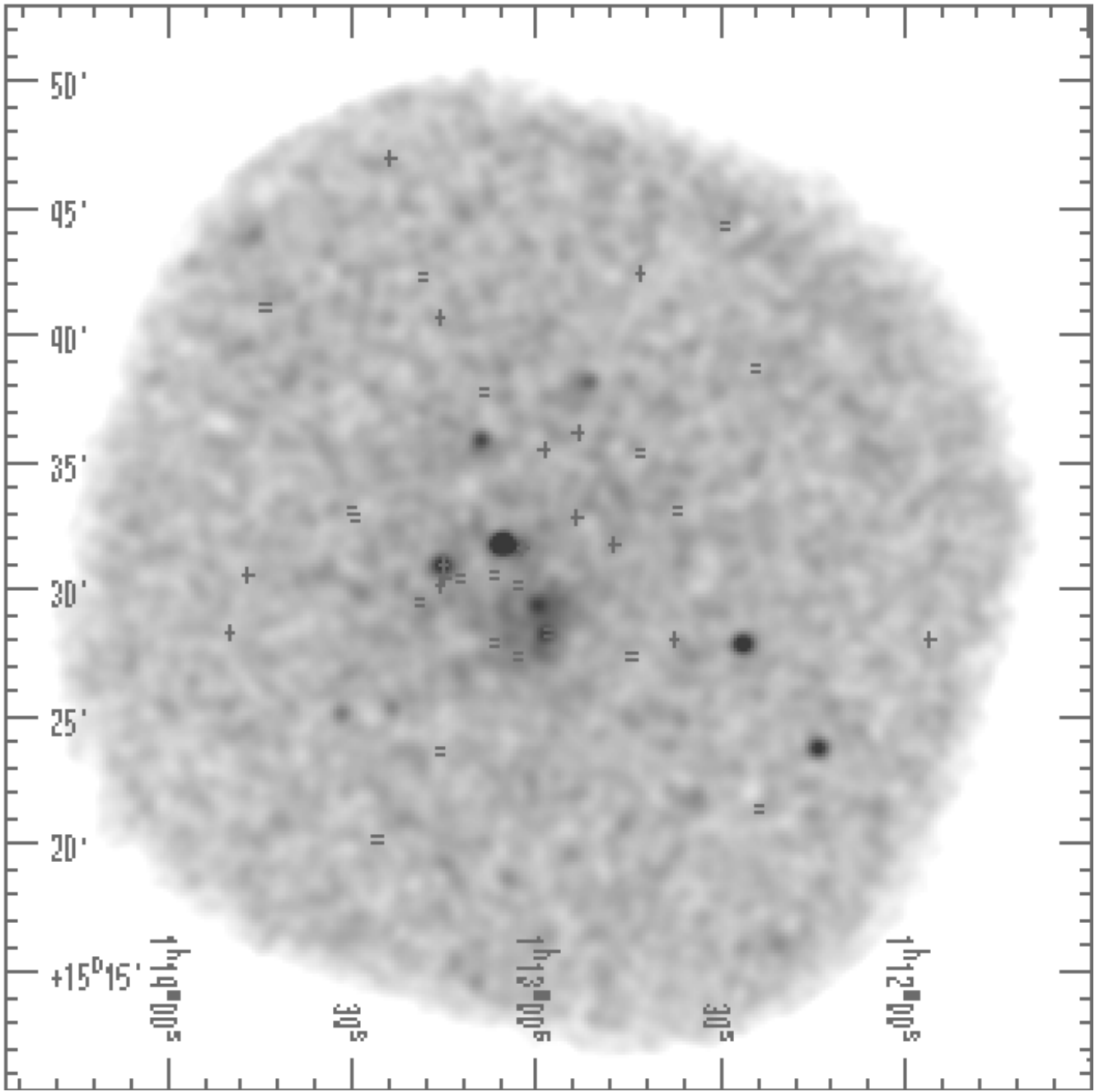,width=17.6cm,angle=0}
 \end{center}
 \caption{Greyscale \emph{ROSAT} HRI image of Abell~160, smoothed using a 
          Gaussian kernel with a dispersion of eight arcseconds.  
          The positions of the 35 cluster galaxies with measured 
          redshifts studied in this paper are marked with symbols 
          such that the brightest 15 are 
          marked with `$+$' and the fainter 20 with `$=$'.  
          These subsamples are discussed later in the paper.}  
 \label{fig:overlay}
\end{figure*}

In order to obtain a more objective measure of the importance of ram
pressure stripping in poor clusters, and the frequency with which it
produces wakes behind galaxies, we need to look at a well-defined sample
of cluster members within a single system.  The cluster Abell~160 provides
an ideal candidate for such a study.  Its richness class of 0 makes it a
typical poor system.  It lies at a redshift of $z=0.045$, and hence at a
distance of $270\,{\rm Mpc}$,\footnote{We adopt a value for the Hubble
constant of $H_{0}=50\,{\rm km}\,{\rm s}^{-1}\,{\rm Mpc}^{-1}$ throughout
this paper.} which is sufficiently close to allow galactic-scale structure
to be resolved in its X-ray emission and corresponds to a size scale of
$\sim79$~kpc~arcmin$^{-1}$.  Furthermore, its Bautz-Morgan class of III
means that it contains quite a number of comparably luminous galaxies, and
one might hope to detect ISM emission most readily from such a sample.  In
addition, its Rood-Sastry classification of C means that its members are
concentrated towards the centre of the cluster:  galaxies lying in the
cluster core, where the ICM density is high, will be most affected by ram
pressure stripping.  Finally, Pinkney (1995) has obtained positions and
redshifts for an almost complete, independently-defined set of galaxies in
the field of Abell~160, providing an objective sample of cluster members
for this study. 

We therefore obtained a deep \emph{ROSAT} HRI X-ray image of Abell~160 in
order to investigate ISM stripping in this typical poor cluster.  We use a
Galactic H\textsc{i} column density towards A160 of
$4.38\times10^{20}$~cm$^{-2}$ (Stark et al. 1992) throughout this paper. 
In the next section, we present the X-ray observation and the redshift
data employed.  Section~3 describes an objective method for detecting and
quantifying wake features in the X-ray data, and Section~4 presents the
results of applying this approach to the Abell~160 data.  We conclude in
Section~5 with a discussion of the interpretation of the results.

% * Table 1 *
\begin{table*}
 \centering
 \caption{\emph{ROSAT}~HRI observations of Abell~160.}
 \begin{tabular}{cccc}      \hline
        {\em Observation Start Date} & {\em Observation End Date} & {\em 
         Number of Used OBIs} & {\em Total Time (s)} \\
        \hline
        1996~Dec~30 & 1997~Jan~19 & 14 & 36965 \\ 
        1997~Jul~01 & 1997~Jul~30 & 8 & 16501 \\ 
        1997~Dec~30 & 1998~Jan~07 & 10 & 16943 \\ 
        \hline  \\
 \end{tabular}
 \protect\label{tab:log}
\end{table*}

% *** Section Two ***
\section{X-Ray data and Optical Redshifts}
\label{sec:data}

Abell~160 was observed with the \emph{ROSAT} HRI in three pointings
(1997~January and July, and 1998~January) for a total integration time of
70.4~ksec (see Table~\ref{tab:log}).  The data were reduced with the
\emph{ROSAT} Standard Analysis pipeline, with subsequent analysis
performed using the IRAF/PROS software package. 
 
Point sources detected in the three X-ray observations were used to
examine the registration in relation to the nominal \emph{ROSAT} pointing
position.  A \emph{Digitized Sky Survey} image of Abell~160 was employed
to provide the optical reference frame.  The second observation set was
shifted $\sim-0.9$ arcseconds east and $\sim1.7$~arcsec south, and the
third set was shifted $\sim0.3$~arcsec east and $\sim1.8$~arcsec south. 
Registration of all three sets of observations was then within
0.5~arcseconds of the optical reference, tied down to five sources.  A
greyscale image of the merged X-ray dataset is shown in
Figure~\ref{fig:overlay}. 

The centroid of the diffuse X-ray emission in this image was
calculated by interpolating over any bright point sources, and
projecting the emission down on to two orthogonal axes.  
Fitting a Gaussian to each of these one-dimensional distributions then 
gives a robust estimate for the centroid of the emission.  
This procedure yielded a location of 
\[ \left. 
\begin{array}{c} 
\alpha_{2000.0}=01^{\mathrm{h}}~13^{\mathrm{m}}~05^{\mathrm{s}} 
\\ 
\delta_{2000.0}=+15^{\circ}~29^{\prime}~48^{\prime\prime} 
\end{array} 
\right\} 
\pm43~\mathrm{arcsec}, \]
which was adopted as the cluster centre for the subsequent analysis.  

\begin{figure}
 \centering
 \psfig{file=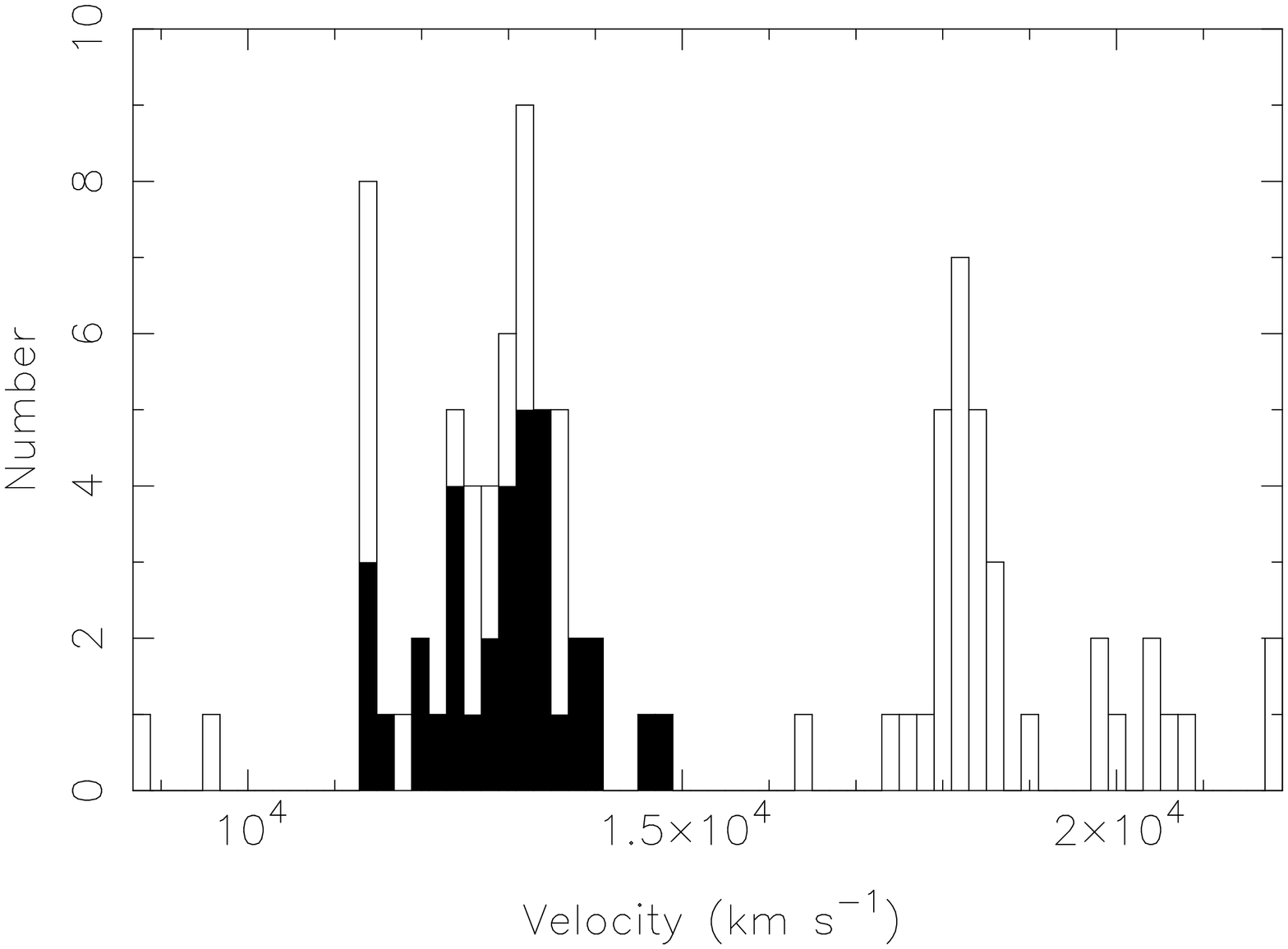,width=8cm,angle=270}
 \caption{Histogram showing the distribution of velocities of galaxies
in the field of Abell~160.  The subsample of galaxies taken 
to be cluster members, and which are in the field of view of 
\emph{ROSAT's} High Resolution Imager, is highlighted as a solid histogram.}  
 \label{fig:veldist}
\end{figure}

Pinkney (1995) obtained redshifts for the 94 brightest galaxies in the
field of A160 using the MX~multi-object spectrograph on the Steward
Observatory 2.3m telescope.  Figure~\ref{fig:veldist} shows the resulting
velocity distribution.  In order to investigate the X-ray properties of
normal cluster members, we have excluded the central galaxy since it
contains a twin-jet radio source (Pinkney 1995), so its X-ray emission may
well contain a significant contribution from the central AGN.  There are
91 galaxies within 8,000~km~s$^{-1}$ of the twin-jet source
($v_{\mathrm{TJ}}=13,173\pm100$~km~s~$^{-1}$); some of these form a
background cluster detected at approximately 18,000~km~s$^{-1}$.  After
further eliminating galaxies outside the field of view of the HRI, we end
up with a sample of 35 cluster members whose X-ray emission we wish to
quantify.  This subsample, highlighted in Figure~\ref{fig:veldist}, has a
line-of-sight velocity dispersion of 560~km~s$^{-1}$, directly comparable
to other poor clusters.  The locations of these galaxies are marked on
Figure~\ref{fig:overlay} where the different symbols indicate different
subsamples of galaxies based upon optical luminosity, as described in the
figure's caption.

% *** Section Three ***
\section{Detecting Wakes}
\label{sec:wakedetect}

Having combined the optical cluster member locations with the X-ray data,
we now turn to trying to see whether there is X-ray emission associated
with any individual galaxy, and whether it takes the form of an X-ray
wake.  On examining Figure~\ref{fig:overlay}, the eye is drawn to a number
of cases where there is an enhancement in the X-ray emission near, but
offset from, the optical galaxy position --- see, for example, the galaxy
at RA~1:12:38, Dec~15:28:03.  It would be tempting to ascribe these
near-coincidences to X-ray wakes.  However, it is also clear from
Figure~\ref{fig:overlay} that there are many apparent enhancements in the
X-ray emission that are totally unrelated to cluster members: some will be
from foreground and background point sources, while others are probably
substructure or noise associated with the ICM itself.  What we need,
therefore, is some objective criterion for assessing the probability that
any given wake is a true association rather than a chance superposition. 
Further, even if we cannot unequivocally decide whether some particular
feature is real, we need to be able to show that there are too many
apparent wakes for all to be coincidences.

% * Figure 2 *

The test adopted to meet these requirements is as follows.  First, for
each galaxy we must seek to detect the most significant wake-like emission
that might be associated with it.  We must therefore choose a range of
radii from the centre of the galaxy in which to search for a wake. 
Scaling the wakes previously detected in other clusters to the distance of
Abell~160, we might expect enhanced X-ray emission at radii $r < 8\,{\rm
arcsec}$, equivalent to a galaxy distance of approximately 10~kpc.  We
also want, as far as possible, to exclude emission from any faint central
AGN component in the galaxy, so we only consider emission from radii $r >
3\,{\rm arcsec}$.  Balsara, Livio \& O'Dea (1994) found wake-like
structure in high resolution, hydrodynamic simulations with a scalelength
of $\sim2$~arcseconds at the distance of Coma (see also Stevens, Acreman
\& Ponman 1999); in the poor environment of Abell~160 we expect wakes to
be longer as their formation should be dominated by ISM stripping.  We
adopt the annulus $3 < r < 8\,{\rm arcsec}$ to search for wakes:  counts
in annuli with larger inner and outer radii were also performed but did
not improve the statistical results described below. 

In our chosen annulus, we search for the most significant emission feature
by taking a wedge with an opening angle of 45 degrees, rotating about the
centre of the galaxy in 10 degree increments, and finding the angle that
produces the maximum number of counts in the intersection of the wedge and
the annulus.  Finally, the contribution to the emission in this wedge from
the surrounding ICM is subtracted by calculating an average local
background between radii $25 < r< 60\,{\rm arcsec}$, centered on the
galaxy and each comparison region at the same cluster radius (see below),
to give a brightest wake flux, $f_{\rm wake}$. 

To provide a diagnostic as to the nature of this wake, its direction on
the plane of the sky, $\Theta$, was also recorded.  This angle was
measured relative to the line joining the galaxy to the cluster centre, so
that $|\Theta| = 0^{\circ}$ corresponds to a wake pointing directly away
from the cluster centre, while $|\Theta|=180^{\circ}$ indicates one
pointing directly toward the cluster centre.\footnote{Under the assumption
of approximately spherical symmetry in the cluster, there is no physical
information in the sign of $\Theta$.}

Having found this strongest wake feature, we must assess its significance. 
To do so, we simply repeated the above procedure using
$n_{\mathrm{comp}}=100$ points for each galaxy at the same projected
distance from the cluster centre, but at randomly-selected azimuthal
angles.  These comparison points were chosen to lie at the same distance
from the cluster centre so that the properties of the ICM and the amount
of vignetting in the \emph{ROSAT} image were directly comparable to that
at the position of the real galaxy.  As noted above, counts in `background
annuli' were also acquired and all these counts were averaged together in
order to obtain a value for the background to be subtracted from the
counts in each wedge region.  Fewer comparison regions were used for the
12 galaxies closest to the cluster centre, as otherwise the count regions
would overlap. 

The comparison regions and real
data were then sorted by their values of $f_{\rm wake}$, from faintest
to brightest, and the rank of the galaxy (i.e. the position of the
real data in this ordered list), $\mathrm{Rank}_{\mathrm{gal}}$,
computed.  
The statistic
\begin{equation} 
k=\frac{\mathrm{Rank}_{\mathrm{gal}}}{n_{\mathrm{comp}}+1} 
\label{eq:frank} 
\end{equation} 
was then calculated.  
Clearly, if all the apparent galaxy wakes were
spurious, then nothing would differentiate these regions from the
comparison regions, and we would expect $k$ to be uniformly
distributed between 0 and 1.  
For significant wake features, on the
other hand, we would expect the distribution of $k$ values to be
skewed toward $k \sim 1$.  

As an additional comparison to the X-ray emission around galaxies in
Abell~160 we performed the same analysis on 70.4~ksec of `blank field'
data, extracted from the \emph{ROSAT} Deep Survey, which encompasses
$\sim1,320$~ksec of HRI pointings towards the Lockman Hole (see e.g. 
Hasinger et al. 1998).  We searched for wake-like features around 35
random positions across this HRI field, using comparison regions for each
`galaxy' as defined above. 

% *** Section Four ***
\section{Results}
\label{sec:results}

\begin{figure}
 \centering
 \psfig{file=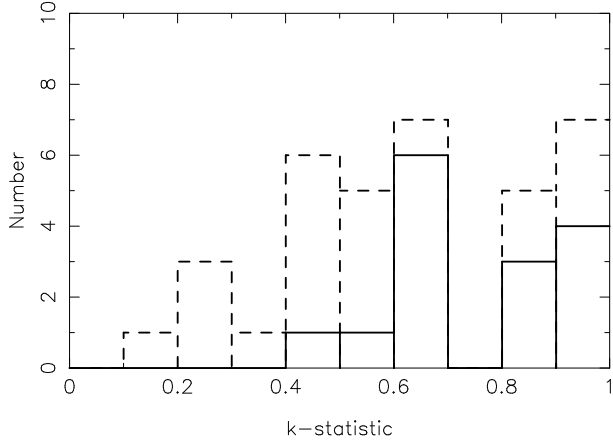,width=8cm,angle=270}
 \caption{Histogram of the values of the $k$ statistic derived for both the
          complete sample of 35 cluster members (dashed line) and for the 
          subsample of the 15 brightest cluster members (solid line).  }
 \label{fig:khist}
\end{figure}

We have applied the above analysis to the confirmed cluster members using
IRAF software packages and the merged QP datafile.  Figure~\ref{fig:khist}
shows the distribution of the $k$ statistic for both the complete sample
of 35 galaxies and the subsample of the 15 brightest galaxies.  The full
sample of A160 galaxies yields a reduced chi-squared value of
$\chi^{2}=1.617$ when fitted by a uniform distribution, which is
approximately a $3\sigma$ deviation;  the probability of obtaining
$\chi^{2}\geq1.617$ for 34 degrees of freedom is only
$\sim5.1\times10^{-3}$.  The full sample has $\left<k\right> = 0.628$ and
the subsample of the 15 brightest galaxies yields $\left<k\right> =
0.741$; indeed, selecting subsamples of the 20 or so optically brightest
galaxies always gives $\left<k\right> > 0.7$ and the distributions are
clearly skewed towards $k=1$.  This suggests that we have detected
significant wake-like excesses in these data. 

% * Figure 3 *

\begin{figure}
 \centering
 \psfig{file=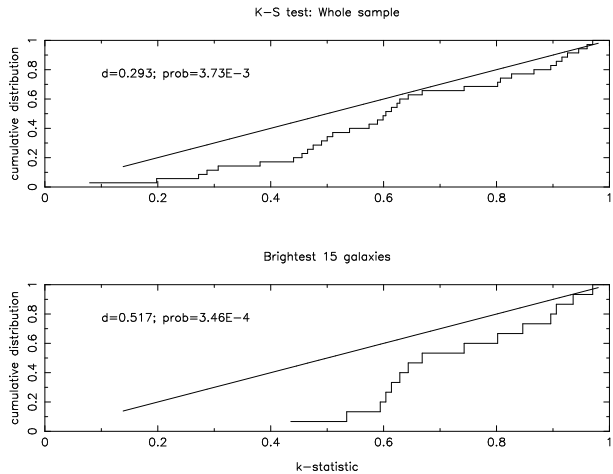,width=8cm,angle=270}
 \caption{Plot of the cumulative distributions of the Kolmogorov-Smirnov 
test for the whole sample of 35 cluster galaxies (top panel) and for the 
subsample of the 15 optically brightest galaxies (bottom panel).  
$d$ and \emph{prob} are explained in the text.}  
 \label{fig:kspanels}
\end{figure}

Kolmogorov-Smirnov tests were performed to compare the $k$ statistic
results to a uniform distribution.  Figure~\ref{fig:kspanels} presents
these test results for the complete sample of galaxies as well as the
brightest 15 subsample.  The figure's annotation gives the values of the
K-S statistic, $d$, which is simply the greatest distance between the
data's cumulative distribution and that of a uniform distribution, and
\emph{prob}, which is a measure of the level of significance of $d$.  The
probability of the detected features arising from chance is less than
$3.5\times10^{-4}$ for the bright sample and $3.8\times10^{-3}$ for the
entire sample of 35 galaxies.  It is clear from Figure~\ref{fig:kspanels}
that the more luminous galaxies do not follow a uniform distribution in
$k$-space. 

% * Figure 4 *

The analysis applied to the Deep Survey `blank field' data yielded a mean
value for the $k$ statistic of $\left<k\right>=0.48$ indicating that the
results are uniformly distributed in $k$-space and that we do not detect
`wakes' in this comparison field.  Indeed, assuming a uniform distribution
for $k$, the fit to the data has a chi-squared value of $\chi^{2}=1.015$,
implying that the values of $k$ are uniformly distributed to high
precision. 

\begin{figure}
 \centering
 \leavevmode
 \epsfxsize 0.9\hsize
 \epsffile{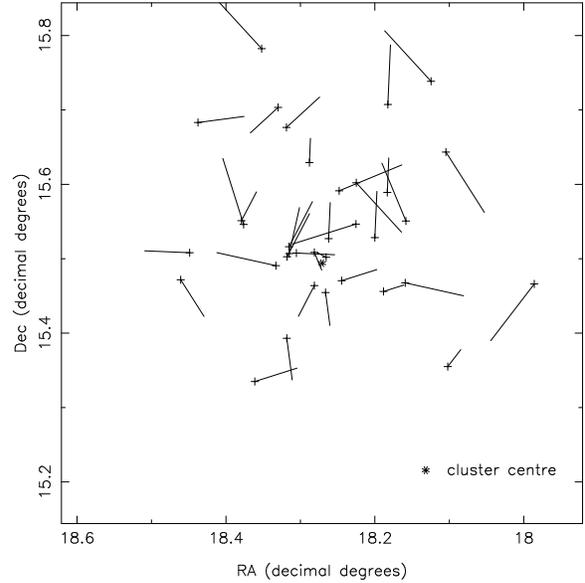}
 \caption{Directions of the wake-like features found for the 35 cluster 
          galaxies in Abell~160.  
          The positions of the galaxies are marked by crosses and the 
          lines represent the wakes.  
          Length of line is proportional to wake strength (as determined 
          by the $k$ statistic).  
          The centre of the cluster is also shown for reference.  }
 \label{fig:directions}
\end{figure}

In order to investigate the nature of the wake-like excesses in the A160
data, we now consider the distribution of their directions on the sky. 
Figure~\ref{fig:directions} shows the directions of the strongest
wake-like features found using the $k$ statistic analysis, for all 35
cluster members.  The lines on this figure represent the wakes and their
lengths are drawn proportional to wake strength. 

% * Figure 5 *

\begin{figure*}
 \centering
 \psfig{file=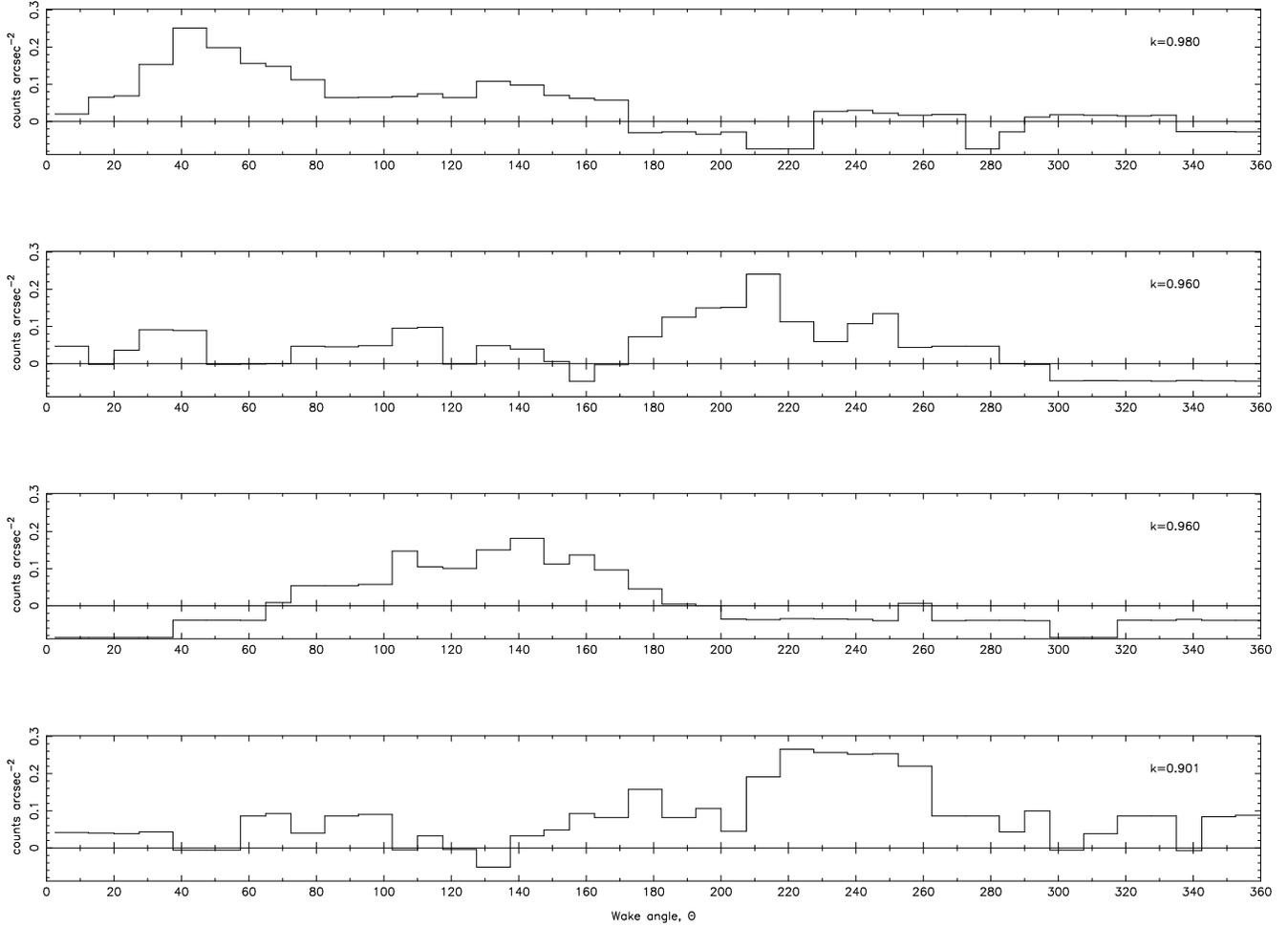,width=17.6cm,angle=270}
 \caption{Azimuthal distributions of the counts around four galaxies for 
which $k>0.9$.  Here the azimuthal angle (with respect to the parent 
galaxy) is defined such that $0^{\circ}$ corresponds to north on the sky 
and the angle increases counter-clockwise.  The values of $k$ for each 
galaxy are noted in the panels.  }
 \label{fig:wprofs}
\end{figure*}

The azimuthal distributions of the counts found in some of the
highest-ranked wakes are shown in Figure~\ref{fig:wprofs}.  The wake
profiles show wakes for these galaxies to be statistically significant
with a mean net count of $\sim5$, corresponding to a mean net X-ray flux
of $1.7\times10^{-15}$erg~cm$^{-2}$~s$^{-1}$.  The wake fluxes translate
into X-ray luminosities in the range $(1-2)\times10^{40}$erg~s$^{-1}$. 
Grebenev et al. (1995) used a wavelet transform analysis to study the
small-scale X-ray structure of the richness class 2 cluster Abell 1367: 
they found 16 extended features of which nine were associated with
galaxies and had luminosities in the range
$(3-30)\times10^{40}$erg~s$^{-1}$.  They concluded that the features could
be associated with small galaxy groups, as suggested by Canizares,
Fabbiano \& Trinchieri (1987), rather than individual galaxies.  The
wake-like features we have detected have X-ray luminosities of the same
order as individual galaxies in Abell 160, and the emission is clearly
confined to extensions in specific directions away from the galaxies. 
Furthermore, the features noted by Canizares, Fabbiano \& Trinchieri
(1987) have size scales $\sim1^{\prime}$, much larger than the expected
wake size at the distance of A1367 and so not directly comparable with the
current work.

% * Figure 6 *

Figure~\ref{fig:scatter} shows the distribution of apparent wake angles as
a function of the galaxies' radii in the cluster, for different strengths
of wake as quantified by the $k$ statistic.  As we might expect, for low
values of $k$ where the wake is almost certainly a noise feature, the
values of $\left|\Theta\right|$ are randomly distributed between
$0^{\circ}$ and $180^{\circ}$.  However, for values of $k>0.7$, which are
unlikely to be attributable to noise, there is only one wake pointed at an
angle of $\left|\Theta\right|>135^\circ$.  If the distribution of wake
directions was intrinsically isotropic, the probability of finding only
one of the 12 most significant wakes in this range of angles is only 0.01. 
Given the \emph{a posteriori} nature of this statistical measure, its high
formal significance should not be over-interpreted.  Nonetheless, there
definitely appears to be a deficit of wakes pointing toward the cluster
centre.

% *** Section Five ***
\section{Discussion}
\label{sec:discussion}

As a first attempt at an objective determination of the frequency of wakes
behind cluster galaxies, we have found significant excesses of X-ray
emission apparently offset from their host galaxies.  Before exploring the
possible astrophysical meaning of such features, we must rule out more
prosaic possibilities. 

If the X-ray emission were truly centred on the galaxies, we would still
detect offset X-ray emission if there were significant positional errors
in the optical galaxy locations.  The uncertainties on these positions,
however, are much less than the radii at which we have detected the wakes,
so this possibility can be excluded. Similarly, an overall mismatch
between the optical and X-ray reference frames would produce offsets
between X-ray and optical locations of coincident sources, but the
distribution of apparent offset directions shown in
Figure~\ref{fig:directions} is not consistent with the coherent pattern
that one would expect from either an offset or a rotation between the two
frames.

% * Figure 7 *
\begin{figure}
 \centering
 \psfig{file=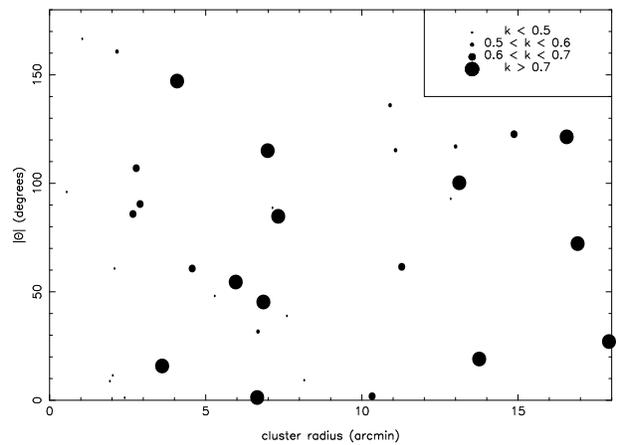,width=8cm,angle=270}
 \caption{Scatter plot of the projected angle
          $\left|\Theta\right|$ of the 35 brightest wedge features as a 
          function of distance from cluster centre.  
          Larger point sizes reflect greater values of the 
          $k$ statistic found for each galaxy, as 
          given in the key to the figure.  }
 \label{fig:scatter}
\end{figure}

We could explain the excess of sources where the X-ray emission lies at
larger radii in the cluster than the optical position if the spatial scale
of the optical data had been underestimated relative to that of the X-ray
data.  But both the \emph{ROSAT} and optical data image scales are
extremely well calibrated.  Further, if such mismatch in magnification
were responsible for the effect, one would expect the radial offsets to
increase with distance from the field centre, and Figure~\ref{fig:scatter}
provides no evidence that the wakes become more radially oriented at large
distances from the cluster centre. 

A further possibility is that the distorted nature of the X-ray emission
could arise from an asymmetry in the \emph{ROSAT} HRI point-spread
function (PSF).  Such asymmetries are documented [see, for example, Morse
et al.\ (1995)], but the observed shape of the HRI PSF actually becomes
tangentially extended at large off-axis angles, so one would expect the
wakes to be oriented at angles of $|\Theta| \sim 90^\circ$.  The three
wakes at very large radii and $|\Theta| \simeq70-120^\circ$ in
Figure~\ref{fig:scatter} could well result from this phenomenon, but there
is no evidence for any such effect at smaller radii. 

We are therefore forced to return to trying to find an astrophysical
explanation for the bulk of the observed wakes.  As discussed in the
introduction, an individual galaxy can emit at X-ray wavelengths due to
both its hot ISM component and its contingent of X-ray binaries. Such
emission could extend to the radii where we have been searching for wakes,
or appear to do so due to the blurring influence of the PSF, so we might
expect some wake-like features to appear simply due to this component. 
Such asymmetric wake features could arise from Poisson noise on
intrinsically symmetric emission, or it could reflect a real asymmetry in
the emission.  For example, the emission from X-ray binaries could be
dominated by one or two ultra-luminous sources in the outskirts of a
galaxy, leading to an offset in the net X-ray emission.  Even the X-ray
wake phenomenon that we are seeking to detect can be described as an
asymmetric distortion in the normal ISM emission.  How, then, are we to
distinguish between these possibilities? 

Perhaps the best clue as to the nature of the detected asymmetric emission
comes from the distribution of the angles at which it is detected,
$\Theta$.  As we have described above, there is a deficit of wakes
pointing toward the cluster centre.  It is hard to see how such a
systematic effect can be attributed to any of the more random processes
such as Poisson noise on a symmetric component, or even the azimuthal
distribution of X-ray binaries within the galaxy.  It therefore seems
highly probable that we are witnessing the more systematic wake phenomenon
that we seek.  If a wake indicates the direction of motion of the galaxy,
then the deficit of detections at large values of $|\Theta|$ implies that
the production mechanism becomes ineffective when a galaxy is travelling
out from the cluster centre. This conclusion has a simple physical
explanation: if a galaxy is travelling on a reasonably eccentric orbit, by
conservation of angular momentum it will spend a large fraction of its
time close to the orbit's apocentre.  During this period, its velocity is
slow and the ICM it encounters is tenuous, so it is able to retain its
ISM.  In fact, continued mass loss from stellar winds and planetary
nebulae means that the amount of gas in its ISM will increase. 
Ultimately, however, its orbit will carry it inward toward the core of the
cluster.  At this point, the galaxy is travelling more rapidly, and
encounters the higher density gas near the centre of the cluster, so ram
pressure stripping becomes more efficient, and a wake of stripped ISM
material will be seen behind the infalling galaxy.  By the time the galaxy
passes the pericentre of its orbit, the ISM will have been stripped away
to the extent that the outgoing galaxy does not contain the raw material
to create a measurable wake, explaining the lack of detected wakes at
large values of $|\Theta|$. 

This simple picture seems to fit the data on Abell~160 rather well; a
similar scenario was invoked by McHardy (1979) to explain the locations of
weak radio sources in clusters.  It is also notable that the beautiful
wake feature behind NGC~1404 in the Fornax Cluster detected by Jones et
al.\ (1997) is oriented such that it points radially away from the cluster
centre.  Clearly, though, more deep X-ray observations of clusters are
required if we are to confirm the widespread applicability of this
scenario.

% ** Acknowledgements **
\subsection*{Acknowledgements}
The authors are grateful to the referee for helpful comments and 
suggestions, and to Ian McHardy for several fruitful discussions.  
ND acknowledges receipt of a PPARC Studentship.  
This research has made use of data obtained from the Leicester Database 
and Archive Service at the Department of Physics and Astronomy, Leicester 
University, UK.  

% *** References ***

\bsp

\end{document}